\documentclass[prb, amsmath,amssymb,reprint,superscriptaddress]{revtex4-1}
\usepackage{graphicx}
\usepackage{natmove}

\begin{document}
\title{Tricritical behavior of two-dimensional intrinsic ferromagnetic semiconducting CrGeTe$_3$}
\author{G. T. Lin}
\affiliation{Key Laboratory of Materials Physics, Institute of Solid State Physics, Chinese Academy of Sciences, Hefei, 230031, China}
\affiliation{University of Science and Technology of China, Hefei, 230026, China}
\author{H. L. Zhuang}
\affiliation{Department of Mechanical and Aerospace Engineering, Princeton University, Princeton, New Jersey 08544, USA}
\author{X. Luo}
\email{xluo@issp.ac.cn}
\affiliation{Key Laboratory of Materials Physics, Institute of Solid State Physics, Chinese Academy of Sciences, Hefei, 230031, China}
\author{B.J. Liu}
\affiliation{University of Science and Technology of China, Hefei, 230026, China}
\affiliation{High Magnetic Field Laboratory, Chinese Academy of Sciences, Hefei, 230031, China}
\author{F. C. Chen}
\affiliation{Key Laboratory of Materials Physics, Institute of Solid State Physics, Chinese Academy of Sciences, Hefei, 230031, China}
\affiliation{University of Science and Technology of China, Hefei, 230026, China}
\author{J. Yan}
\affiliation{Key Laboratory of Materials Physics, Institute of Solid State Physics, Chinese Academy of Sciences, Hefei, 230031, China}
\affiliation{University of Science and Technology of China, Hefei, 230026, China}
\author{Y. Sun}
\affiliation{University of Science and Technology of China, Hefei, 230026, China}
\affiliation{High Magnetic Field Laboratory, Chinese Academy of Sciences, Hefei, 230031, China}
\author{J. Zhou}
\affiliation{MIIT Key Laboratory of Critical Materials Technology for New Energy Conversion and Storage, School of Chemistry and Chemical Engineering,
Harbin Institute of Technology, Harbin 1500001, China}
\author{W. J. Lu}
\affiliation{Key Laboratory of Materials Physics, Institute of Solid State Physics, Chinese Academy of Sciences, Hefei, 230031, China}
\author{P. Tong}
\affiliation{Key Laboratory of Materials Physics, Institute of Solid State Physics, Chinese Academy of Sciences, Hefei, 230031, China}
\author{Z. G. Sheng}
\affiliation{High Magnetic Field Laboratory, Chinese Academy of Sciences, Hefei, 230031, China}
\affiliation{Collaborative Innovation Center of Advanced Microstructures, Nanjing University, Nanjing, 210093, China}
\author{Z. Qu}
\affiliation{High Magnetic Field Laboratory, Chinese Academy of Sciences, Hefei, 230031, China}
\author{W. H. Song}
\affiliation{Key Laboratory of Materials Physics, Institute of Solid State Physics, Chinese Academy of Sciences, Hefei, 230031, China}
\author{X. B. Zhu}
\affiliation{Key Laboratory of Materials Physics, Institute of Solid State Physics, Chinese Academy of Sciences, Hefei, 230031, China}
\author{Y. P. Sun}
\email{ypsun@issp.ac.cn}
\affiliation{High Magnetic Field Laboratory, Chinese Academy of Sciences, Hefei, 230031, China}
\affiliation{Key Laboratory of Materials Physics, Institute of Solid State Physics, Chinese Academy of Sciences, Hefei, 230031, China}
\affiliation{Collaborative Innovation Center of Advanced Microstructures, Nanjing University, Nanjing, 210093, China}
\date{\today}

\begin{abstract}
CrGeTe$_3$ recently emerges as a new two-dimensional (2D) ferromagnetic semiconductor that is promising for spintronic device applications. Unlike CrSiTe$_3$ whose magnetism can be understood using the 2D-Ising model, CrGeTe$_3$ exhibits a smaller van der Waals gap and larger cleavage energy, which could lead to a transition of magnetic mechanism from 2D to 3D. To confirm this speculation, we investigate the critical behavior of CrGeTe$_3$ around the second-order paramagnetic-ferromagnetic phase transition. We obtain the critical exponents estimated by several common experimental techniques including the modified Arrott plot, Kouvel-Fisher method and critical isotherm analysis, which show that the magnetism of CrGeTe$_3$ follows the tricritical mean-field model with the critical exponents $\beta$, $\gamma$, and $\delta$ of 0.240$\pm$0.006, 1.000$\pm$0.005, and 5.070$\pm$0.006, respectively, at the Curie temperature of 67.9 K. We therefore suggest that the magnetic phase transition from 2D to 3D for CrGeTe$_3$  should locate near a tricritical point. Our experiment provides a direct demonstration of the applicability of the tricritical mean-field model to a 2D ferromagnetic semiconductor.
\end{abstract}
\maketitle


\section{Introduction}

Since the successful exfoliation of single layer graphene, 2D materials have been attracting significant interest due to their highly tunable physical properties and immense potential in scalable device applications \cite{novoselov2004electric,novoselov2005two,zhang2005experimental,geim2007rise,RevModPhys.81.109}. However, pristine graphene exhibits no band gap and its inherent inversion symmetry suppresses the spin-orbit coupling (SOC) \cite{zibouche2014transition, PhysRevB.92.144404, RevModPhys.76.323, macdonald2005ferromagnetic, dietl2010ten, deng2011li}. The weak SOC and zero band gap eliminate graphene as a potential candidate for being applied to spintronic devices, which require one to search for alternative 2D materials that extend beyond graphene to other layered materials with van der Waals gaps \cite{PhysRevB.92.144404, RevModPhys.76.323, macdonald2005ferromagnetic, dietl2010ten, deng2011li}. For example, in single-layer MoS$_2$, the large SOC leads to a unique spin-valley coupling which may be useful for spintronic applications \cite{PhysRevB.84.153402,PhysRevLett.108.196802,PhysRevB.86.165108,PhysRevB.88.125301,xu2014spin}. Whereas spintronic devices using 2D materials are still in their infancy \cite{han2014graphene,ohno2000electric,chang2013experimental,PhysRevLett.113.137201}, which is due to the lack of long-range ferromagnetic order that is crucial for macroscopic magnetic effects \cite{PhysRevLett.114.016603,gonzalez2016atomic}. The emergence of ferromagnetism in 2D materials in combination with their rich electrical and optical properties could open up ample opportunities for 2D magnetic, magneto-electric, and magneto-optic applications \cite{ohno2000electric,chang2013experimental,gong2017discovery}.

Recently, Chromium Tellurides Cr$X$Te$_3$ ($X$ = Si, Ge, and Sn) with the centrosymmetric have arisen significant attention because they belong to a rare category of ferromagnetic semiconductors possessing a 2D layered structure\cite{PhysRevB.92.144404,gong2017discovery,PhysRevB.91.235425,carteaux1995crystallographic,ji2013ferromagnetic,tian2016magneto,PhysRevB.92.035407,siberchicot1996band,PhysRevX.3.031002,li2014crxte,PhysRevB.92.165418,PhysRevLett.117.257201,liu2016critical,carteaux19952d}. Extensive theoretical and experiment efforts have been extended towards understanding the properties of these 2D magnets. On the theoretical side, recent studies on Cr$X$Te$_3$ have been focusing on their electronic structure and magnetic properties, particularly predictions of the single-layer properties \cite{PhysRevB.91.235425,PhysRevB.92.035407,siberchicot1996band,PhysRevX.3.031002,li2014crxte,PhysRevB.92.165418,PhysRevLett.117.257201}. On the experimental side, CrSiTe$_3$ and CrGeTe$_3$ have been prepared and characterized \cite{PhysRevB.92.144404,gong2017discovery,carteaux1995crystallographic,ji2013ferromagnetic,tian2016magneto,liu2016critical,carteaux19952d}. Comparing with  CrSiTe$_3$, showing characteristics of a 2D-Ising behavior\cite{PhysRevB.92.144404,liu2016critical,carteaux19952d}, the smaller van der Waals gap and the larger in-plane nearest-neighbor Cr-Cr distance in CrGeTe$_3$  enhance the Curie temperature from 32 K for the  CrSiTe$_3$ to 61 K for the CrGeTe$_3$ \cite{PhysRevB.92.144404,carteaux1995crystallographic,PhysRevB.92.035407,li2014crxte}. In addition, theoretical investigations have suggested that the single-layer CrGeTe$_3$ presents characteristics of 2D-Ising behavior similar to CrSiTe$_3$\cite{li2014crxte,PhysRevLett.117.257201}. By contrast, a scanning magneto-optic Kerr microscopy experiment, single-layer CrGeTe$_3$ represents a close-to-ideal 2D Heisenberg ferromagnetic system using the rigorous renormalized spin wave theory analysis and calculations\cite{gong2017discovery}. It is known that, with the increase of the $X$ atom radius, Cr$X$Te$_3$ presents the smaller van der Waals gap and the larger cleavage energy \cite{PhysRevB.92.144404,carteaux1995crystallographic,PhysRevB.92.035407,li2014crxte}. We suppose that Cr$X$Te$_3$ system may undergo a three dimensional (3D) magnetic phase transition from 2D with the increase of the $X$ atom radius. Therefore, a method to rapidly characterize the critical behavior of single-crystalline CrGeTe$_3$ is crucial. For this purpose, we present a detailed investigation of the critical phenomena of CrGeTe$_3$ using the initial isothermal $M(H)$ curves around the Curie temperature $T_\mathrm{C}$. We find that the critical exponents of CrGeTe$_3$ satisfy the universality class of the tricritical mean-field theory. This indicates that the magnetic phase transition of CrGeTe$_3$ should be close to a tricritical point from 2D to 3D.

\section{Methods}
\label{sec:methods}
Samples of single-crystalline CrGeTe$_3$ were prepared by the self-flux technique \cite{ji2013ferromagnetic}. The XRD data indicated that the powders are single phase with the rhombohedral structure (see Supporting Information). We measured the heat capacity using the Quantum Design physical properties measurement system (PPMS-9T) and characterized the magnetic properties by the magnetic property measurement system (MPMS-XL5). Density functional theory (DFT) calculations were performed using Vienna {\it Ab-initio} Simulation Package \cite{Ref36}. We used the local density approximation \cite{Ref37, Ref38} to treat the electron-electron exchange-correlation interactions. The electron-ion interactions are described by the potentials based on the projector augmented wave method \cite{Ref39, Ref40}.
\section{Results}
\begin{figure}[t]
  \includegraphics[width=8cm]{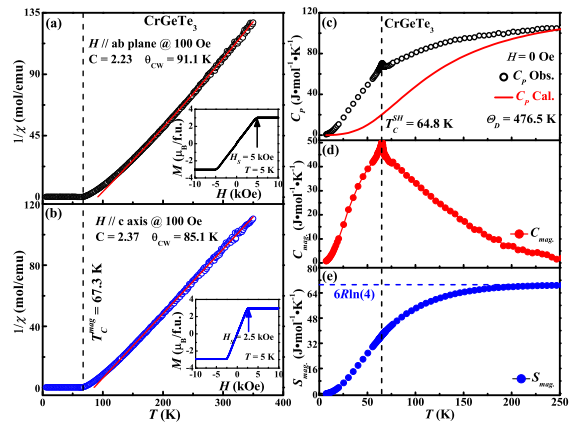}
  \caption{(a) and (b) Temperature-dependent inverse susceptibility 1/$\chi(T)$ of CrGeTe$_3$ under field cooled cooling with an applied magnetic field $H$ of 100 Oe, parallel to the $ab$ plane and $c$ axis, respectively. The red solid lines are the fitted results according to the Curie-Weiss law. The insets show the isothermal magnetization curves $M$($H$) at 5 K. (c) Specific heat $C_p$ as a function of $T$ for CrGeTe$_3$ and the fitted $C_V^\mathrm{Debye}(T)$ using Eqs.\ref{eq:1} and 2; Temperature-dependent magnetic (d) specific heat $C_\mathrm{mag}(T)$ and (e) entropy $S_\mathrm{mag}(T\rightarrow\infty)$. The blue dashed line refers to $S_\mathrm{mag}(T)$ calculated with the magnetic moment S of Cr$^{3+}$ being 3/2.}
  \label{fig:1}
\end{figure}

Figure~\ref{fig:1}(a) and (b) show the temperature-dependent inverse susceptibility 1/$\chi(T)$ of CrGeTe$_3$ under field cooled cooling with applied magnetic field $H$ = 100 Oe, parallel to the $ab$ plane and $c$ axis, respectively. We observe a paramagnetic-ferromagnetic (PM-FM) transition that occurs at a critical temperature of 67.3 K, as determined by the derivative of the susceptibility. This temperature is consistent with the values of 61 K or 70 K reported previously \cite{carteaux1995crystallographic,ji2013ferromagnetic,tian2016magneto}. For a FM system, the 1/$\chi(T)$ above can be described by the Curie-Weiss law resulting from the mean-field theory \cite{Ref41}. The red curves showing the Curie-Weiss law are obeyed only at high temperatures. A close observation of Fig.\ref{fig:1}(a) and (b) reveals that the curves deviate from straight lines at around 150 K, which is much higher than $T_C^\mathrm{mag}$, indicating strong short-range FM spin interactions in CrGeTe$_3$ above $T_C^\mathrm{mag}$. The effective magnetic moment $\mu_\mathrm{eff}$ is determined to be around 4.22$\mu_\mathrm{B}$  (parallel the ab plane) and 4.35$\mu_\mathrm{B}$ (parallel the c axis), which are close to the theoretical value expected for Cr$^{3+}$ of 3.87$\mu_\mathrm{B}$ \cite{carteaux1995crystallographic}. The insets of Fig.\ref{fig:1}(a) and (b) show the isothermal magnetization $M(H)$ at 5 K exhibiting a typical FM behavior with the saturation field $H_S$ of about 5 kOe (parallel to the ab plane) and 2.5 kOe (parallel to the c axis). In addition, the $M(H)$ curves show almost no coercive force for CrGeTe$_3$.

Figure~\ref{fig:1}(c) shows the variation of the zero-field specific heat (SH) $C_p(T)$ with temperature. The sharp anomaly in $C_p(T)$ at 64.8 K corresponds to the Curie temperature $T_C^\mathrm{SH}$. Since CrGeTe$_3$ is a seminconductor \cite{carteaux1995crystallographic}, the electronic contribution to the heat capacity is not considered. The $C_\mathrm{mag}$ can be calculated by the following equations \cite{Ref41}:
\begin{equation}
  \label{eq:1}
C_\mathrm{mag}(T) =C_p(T)-NC_V^\mathrm{Debye}(T)
\end{equation}
and
\begin{equation}
  \label{eq:2}
C_V^\mathrm{Debye}(T) = 9R\bigg(\frac{T}{\Theta_\mathrm{D}}\bigg)^3\int_{0}^{\Theta_\mathrm{D}/T} \frac{x^4e^x}{(e^x-1)^2} dx,
\end{equation}
where $R$ is the molar gas constant, $\Theta_\mathrm{D}$ is the Debye temperature, and $N$ = 5 is the number of atoms per formula unit. The sum of Debye functions accounts for the lattice contribution to the specific heat. We can extract the magnetic contribution $C_\mathrm{mag}(T)$ from the measured specific heat of CrGeTe$_3$. The fitted $C_p(T)$  for CrGeTe$_3$ by Eqs.\ref{eq:1} and 2 over the temperature range from about 7 to 250 K is shown by the red curve in Fig.\ref{fig:1}(c) using the Debye temperature  $\Theta_\mathrm{D}$ = 476.5 K. We observe a sharp peak at $T_C^\mathrm{SH}$ of 64.8 K and there is strong dynamic short-range FM spin interactions above $T_C^\mathrm{SH}$  (see Fig.\ref{fig:1}(d)). The magnetic entropy $S_\mathrm{mag}(T)$ is calculated by
\begin{equation}
  \label{eq:3}
S_\mathrm{mag}(T) =\int_{0}^{T}\frac{C_\mathrm{mag}(T)}{T}dT.
\end{equation}
Fig.\ref{fig:1}(e)  shows the temperature dependence of $S_\mathrm{mag}(T)$. The entropy of CrGeTe$_3$ per mole with completely disordered spins S is
\begin{equation}
  \label{eq:4}
S_\mathrm{mag}(T\rightarrow\infty)=6R\mathrm{ln}(2\mathrm{S}+1).
\end{equation}
Using S = 3/2 for Cr$^{3+}$, we obtain $S_\mathrm{mag}(T\rightarrow\infty)$ of 69.2 J/(mol$\cdot$K). However, we observe the $S_\mathrm{mag}$ is 64.7 J/(mol$\cdot$K) at 150 K in Fig.\ref{fig:1}(e), which is smaller than $S_\mathrm{mag}(T\rightarrow\infty)$.  Note that there is an error of about 10\% \cite{PhysRevB.72.174404} in our measurement due to the fitting of the optical phonon contributions at high temperatures. In spite of this small error, our result indicates the strong short-range FM spin interactions above $T_C^\mathrm{SH}$.

\begin{figure}[t]
  \includegraphics[width=8.5cm]{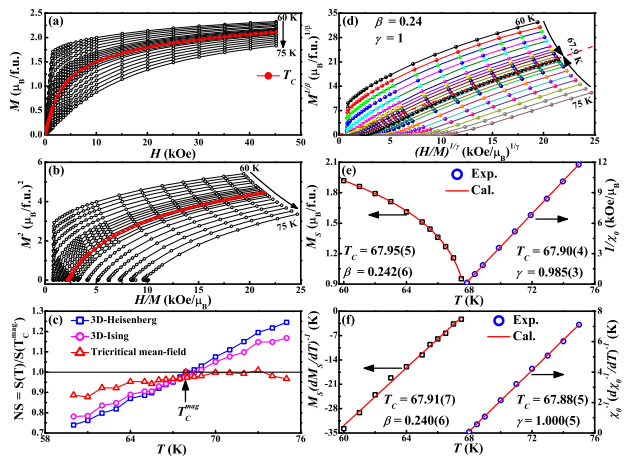}
  \caption{(a) Initial magnetization of CrGeTe$_3$ around $T_\mathrm{C}$; (b) Arrott plots of $M^2$ versus $H/M$ (the $M(H)$ curves are measured at temperature intervals of 1 K and 0.5 K approaching $T_\mathrm{C}$); (c) Normalized slopes as a function of temperature; (d) Modified Arrott plot ($M^{1/\beta}$ versus $(H/M)^{1/\gamma}$) of isotherms with $\beta$ = 0.24 and $\gamma$ = 1 for CrGeTe$_3$. The red dashed line is the linear fit of isotherm at 67.9 K; (e) Temperature dependence of $M_S$ and $\chi_0^{-1}$. The $T_\mathrm{C}$ and critical exponents are obtained from the fitting of Eqs.S1 and S2; (f) The Kouvel-Fisher plot. The  $T_\mathrm{C}$ and critical exponents are obtained from the linear fit.}
  \label{fig:2}
\end{figure}

As mentioned above, with the $X$ atom radius increases, the Cr$X$Te$_3$ compounds present the smaller van der Waals gap and the larger cleavage energy \cite{PhysRevB.92.144404,carteaux1995crystallographic,PhysRevB.92.035407,li2014crxte}, which may induce a 3D magnetic phase transition. For the purpose of confirmation, we performed a detailed characterization of the critical phenomena using the initial isothermal $M(H)$ curves around $T_\mathrm{C}$ for the CrGeTe$_3$, which are shown in Fig.\ref{fig:2}(a). In the mean-field theory, the critical exponents and critical temperature can be determined from the Arrott plot with $\beta$ of 0.5 and $\gamma$ of 1.0 \cite{Ref49, Ref50}. According to this method, the $M^2$ versus $H/M$ (shown in Fig.\ref{fig:2}(b)) should be a series of parallel straight lines in the higher field range around $T_\mathrm{C}$ and the line at $T$ = $T_\mathrm{C}$ should pass through the origin. Note that the lower-field data mainly represent the arrangement of magnetic domains, which should be excluded from the fitting process \cite{Ref51}. However, all the curves in Fig.\ref{fig:2}(b) show nonlinear behaviors having downward curvature even at high fields, which indicates an non-mean-field-like behavior. Moreover, the positive slope reveals a second-order phase transition according to the criterion proposed by Banerjee \cite{Ref52}. As such, a modified Arrott plot should be employed to obtain the critical exponents.

To determine an accurate model, we obtain a modified Arrott plot following Eq.S5 for single-crystalline CrGeTe$_3$ at different temperatures. Three groups of possible exponents belonging to the 3D Heisenberg model ($\beta$ = 0.365, $\gamma$ = 1.386), 3D-Ising model ($\beta$ = 0.325, $\gamma$ = 1.24) and tricritical mean-field model ($\beta$ = 0.25, $\gamma$ = 1.0) exhibit nearly straight lines in the high field region \cite{Ref45, Ref53}. We calculate their normalized slopes (NS) defined as NS =$ S(T)/S(T_C^\mathrm{mag}$ = 67.3 K). By comparing NS with the ideal value of unity, one can identify the most suitable model \cite{Ref45, Ref53}. Fig.\ref{fig:2}(c) shows the plots of NS versus $T$ employing the three different models, revealing that the tricritical mean-field model is the most appropriate to describe the critical behavior of CrGeTe$_3$.

By proper selections of $\beta$ and $\gamma$, one can clearly show the isotherms are a set of parallel straight lines at high fields as displayed in the Fig.\ref{fig:2}(d). The linear extrapolation from the high-field region gives the spontaneous magnetization $M_S(T,0)$ and the initial  inverse susceptibility $\chi_0^{-1}(T,0)$ (see Fig.\ref{fig:2}(e)) corresponding to the intercepts on the $M^{1/\beta}$ and $(H/M)^{1/\gamma}$ axes, respectively. By fitting the data of $M_S(T,0)$ and to Eqs.S1 and S2, one obtains two new values of $\beta$ = 0.242$\pm$0.006 with $T_\mathrm{C}$ = 67.95$\pm$0.01 and $\gamma$ = 0.985$\pm$0.009 with $T_\mathrm{C}$  = 67.90$\pm$0.09. These results are again very close to the critical exponents of tricritical mean-field model. In addition, these critical exponents and $T_\mathrm{C}$ can be obtained more accurately from the Kouvel-Fisher (KF) method \cite{Ref54}. Hence, one can find that the temperature dependence of $M_S(dM_S/dT)^{-1}$  and $\chi_0^{-1}(d\chi_0^{-1}/dT)^{-1}$  should be straight lines with slopes 1/$\beta$ and 1/$\gamma$, respectively. As seen in Fig.\ref{fig:2}(f), the linear fit yields the $\beta$ of 0.240$\pm$0.006 with $T_\mathrm{C}$ of 67.91$\pm$0.07 and $\gamma$ of 1.000$\pm$0.005 with $T_\mathrm{C}$ of 67.88$\pm$0.05, respectively. Remarkably, the obtained values of the critical exponents and $T_\mathrm{C}$ using the KF method are in excellent agreement with those using the modified Arrott plot based on the tricritical mean-field model. This suggests that the estimated values are self-consistent and unambiguous.

\begin{figure}
  \includegraphics[width = 8cm]{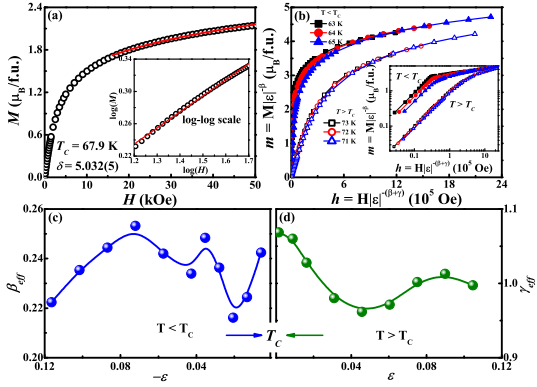}
  \caption{(a) Isothermal $M(H)$ at $T_\mathrm{C}$. The inset shows the alternative plot on a log-log scale and the straight line is the linear fit following Eq.S3; (b) Renormalized magnetization $m$ versus renormalized field $h$ at several typical temperatures around the $T_\mathrm{C}$. The inset shows an alternative plot on a log-log scale; the effective exponents (c) below $T_\mathrm{C}$ and (d)  above $T_\mathrm{C}$ as a function of the reduced temperature $\varepsilon$.}
  \label{fig:3}
\end{figure}

To further validate the above critical exponents $\beta$ and $\gamma$, we study the relation among these exponents. According to Eq.S3, $\delta$ can be directly estimated from the critical isotherm at $T_\mathrm{C}$. Figure \ref{fig:3}(a) shows the isothermal magnetization $M(H)$ at $T_\mathrm{C}$ = 67.9 K. The inset of the same plot has been demonstrated on a log-log scale. The solid straight line with a slope 1/$\delta$ is the fitted result using Eq.S3. From the linear fit we obtained the third critical exponent $\delta$ of 5.032$\pm$0.005. Moreover, the exponent $\delta$ can be calculated by the Widom scaling relation \cite{Ref55,Ref56}

\begin{equation}
  \label{eq:5}
\delta =1+\gamma/\beta.
\end{equation}
Based on the $\beta$ and $\gamma$ values calculated in  Fig.\ref{fig:2}(e) and (f), Eq.\ref{eq:5} yields $\delta$ of 5.070 $\pm$ 0.006 and 5.167 $\pm$ 0.006, respectively. We emphasize that these values are very close to the results from the experimental critical isothermal analysis. Therefore, the critical exponents obtained in this study basically obey the Widom scaling relation, showing that the obtained $\beta$, $\gamma$ and $\delta$ are reliable.

Finally, these critical exponents should follow the scaling equation (Eq.S6) in the critical region. The scaling equation indicates that $m$ versus $h$ forms two universal curves for $T > T_\mathrm{C}$ and  $T < T_\mathrm{C}$, respectively. Based on Eq.S7, the isothermal magnetization around the critical temperatures for CrGeTe$_3$ has been plotted in Fig.\ref{fig:3}(b). All experimental data in the higher-field region collapse onto two universal curves, in agreement with the scaling theory. The inset of Fig.\ref{fig:3}(b) shows the corresponding log-log plot. Similarly, all the points  collapse into two different curves in the higher-field region. This result shows again that the obtained results of the critical exponents and $T_\mathrm{C}$ are valid.

To further examine the convergence of the critical exponents, the effective exponents $\beta_\mathrm{eff}$ and $\gamma_\mathrm{eff}$ can be are obtained by Eqs.S8 and S9 for CrGeTe$_3$. As shown in Fig.\ref{fig:3}(c) and (d), both $\beta_\mathrm{eff}$ and $\gamma_\mathrm{eff}$ show a non-monotonic variation with $\varepsilon$ (see Eq.S4). The lowest $\varepsilon$ ($\varepsilon_\mathrm{min}$) are 5.89$\times$10$^{-3}$ and 1.47$\times$10$^{-3}$ for $\beta_\mathrm{eff}$ and $\gamma_\mathrm{eff}$, respectively. We obtain the effective exponents $\beta_\mathrm{eff}$ of 0.242 and $\gamma_\mathrm{eff}$ of 1.069, indicating that both $\beta_\mathrm{eff}$ and $\gamma_\mathrm{eff}$ are converged when the temperature approaches $T_\mathrm{C}$.

The experimental critical exponents of CrGeTe$_3$, as well as the theoretical values of CrSiTe$_3$, MnSi and some other manganites based on various models, are summarized in Table~\ref{table:1}. It is seen that the critical exponents for MnSi and doped manganites are consistent with those of tricritical mean-field theory \cite{Ref45,Ref53,Ref60,Ref61}. These compounds have the same characteristics,{\it i.e.}, a tricritical point separating the first-order from the second-order ferromagnetic phase transitions. This phenomenon shows that the element substitution \cite{Ref53,Ref61}, hole or electric doping \cite{Ref60}, and external magnetic field\cite{Ref45} can induce the tricritical behavior. However, CrGeTe$_3$ presents a second-order ferromagnetic phase transitions\cite{PhysRevB.92.144404,carteaux1995crystallographic,ji2013ferromagnetic,tian2016magneto,PhysRevB.92.035407} and our results indicate that the critical behavior of CrGeTe$_3$ is close to the theoretical value of tricritical mean-field model. Comparing with CrSiTe$_3$, showing characteristic of 2D-Ising model \cite{PhysRevB.92.144404,liu2016critical,carteaux19952d}, the smaller van der Waals gap and the larger planar nearest-neighbor Cr-Cr distance of CrGeTe$_3$ enhances the Curie temperature from 32 K for the CrSiTe$_3$ to 61 K for the CrGeTe$_3$ \cite{PhysRevB.92.144404,carteaux1995crystallographic,ji2013ferromagnetic,tian2016magneto,PhysRevB.92.035407}. In addition, the neutron scattering and isothermal magnetization experiments yield a critical exponent $\beta$ of around 0.151 or 0.17 for CrSiTe$_3$ \cite{PhysRevB.92.144404,liu2016critical,carteaux19952d}, which is close to the value expected for a 2D transition ($\beta_\mathrm{2D}^\mathrm{Ising}$= 0.125) and well below the values expected for a 3D transition ($\beta_\mathrm{2D}^\mathrm{Ising}$ = 0.326). Our results yield a critical exponent $\beta$ of 0.24 for CrGeTe$_3$ that is close to the critical exponent of the tricritical mean-field model. Hence, the increase of the $X$ atom radius, facilitating super exchange coupling between the Cr atoms via the Te atom and leading to the smaller van der Waals gap in Cr$X$Te$_3$ system \cite{PhysRevB.92.144404,carteaux1995crystallographic,ji2013ferromagnetic,tian2016magneto,PhysRevB.92.035407}, could induce a tricritical magnetic phase transition in the CrGeTe$_3$ single crystal.

\begin{table*}[tb]
  \caption{Critical exponents of CrGeTe$_3$ with various theoretical models, CrSiTe$_3$ and other related materials with tricritical mean-field model (SC = single crystal; PC = polycrystalline; cal = calculated from Eq.\ref{eq:5}).}
  \begin{ruledtabular}
    \begin{center}
      \begin{tabular}{ccccccc}
       Composition & Referecne & $T_\mathrm{C}(\mathrm{K})$ & Technique & $\beta$ & $\gamma$ & $\delta$\\
        \hline
       CrGeTe$_3^\mathrm{SC}$ & This work & 67.9 & Modified Arrott plot &0.242$\pm$0.006 &0.985$\pm$0.003&5.070$\pm$0.006$^\mathrm{cal}$\\
                                                   &                 &         & Kouvel-Fisher method &0.240$\pm$0.006 &1.000$\pm$0.005&5.167$\pm$0.006\\
                                                   &                 &         & Critical isotherm & & &5.032$\pm$0.005$^\mathrm{cal}$\\
         Tricritical mean-field                                         &  [\onlinecite{Ref52}]      &         & Theory &0.25 & 1 &5\\
Mean-field                                         &  [\onlinecite{Ref49}][\onlinecite{Ref50}]      &         & Theory &0.5 & 1 &3\\
3D-Heisenberg theory                                         &  [\onlinecite{Ref49}][\onlinecite{Ref50}]      &         & Theory &0.365 & 1.386 &4.8\\
3D-Ising                                        &  [\onlinecite{Ref49}][\onlinecite{Ref50}]      &         & Theory &0.325 & 1.24 &4.82\\
       CrSiTe$_3^\mathrm{SC}$ & [\onlinecite{liu2016critical}]  & 31 & Modified Arrott plot &0.170$\pm$0.008 &1.532$\pm$0.001&10.012$\pm$0.047$^\mathrm{cal}$\\
       MnSi$^\mathrm{SC}$ & [\onlinecite{Ref45}]  & 30.5 & Modified Arrott plot &0.242$\pm$0.006 &0.915$\pm$0.003&4.734$\pm$0.006\\
La$_{0.1}$Nd$_{0.6}$Sr$_{0.3}$MnO$_3$$^\mathrm{PC}$ & [\onlinecite{Ref53}]  & 249.3 & Modified Arrott plot &0.257$\pm$0.005 &1.12$\pm$0.03&5.17$\pm$0.02\\
La$_{0.9}$Te$_{0.1}$MnO$_3$$^\mathrm{PC}$ & [\onlinecite{Ref60}]  & 239.5 & Modified Arrott plot &0.201$\pm$0.003 &1.27$\pm$0.04&7.14$\pm$0.04\\
La$_{0.6}$Ca$_{0.4}$MnO$_3$$^\mathrm{PC}$ & [\onlinecite{Ref61}]  & 265.5 & Modified Arrott plot &0.25$\pm$0.03 &1.03$\pm$0.05&5.0$\pm$0.8\\

      \end{tabular}
    \end{center}
  \end{ruledtabular}
  \label{table:1}
\end{table*}

\begin{figure}[t]
\center
  \includegraphics[width=8.5cm]{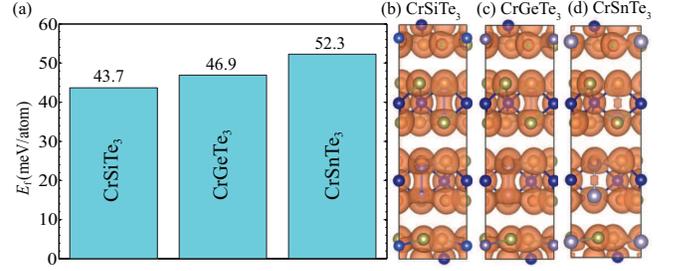}
  \caption{(a) Formation energy of single-layer Cr$X$Te$_3$; The formation energy of single-layer CrSiTe$_3$ is adopted from Ref.[\onlinecite{Ref62}] (b), (c), and (d) charge density of bulk Cr$X$Te$_3$ with an isosurface value of 0.05 $e$/$r_\mathrm{Bohr}^3$.}
  \label{fig:4}
\end{figure}

Although single-crystalline CrSnTe$_3$ has not yet been synthesized, we speculate that the magnetism of CrSnTe$_3$ should be closer to the 3D-Ising model. To support this assumption, we perform DFT calculations with the same calculation parameters that were used in Ref.[\onlinecite{Ref62}]. Figure~\ref{fig:4}(a) shows the calculated formation energy $E_\mathrm{f}$, which is defined as the energy cost of extracting a sheet of single-layer Cr$X$Te$_3$ from their bulk counterparts. As can be seen, $E_\mathrm{f}$ increases as the species vary from Si to Ge. This is consistent with the larger theoretical cleavage energy of single-layer CrGeTe$_3$ than that of CrSiTe$_3$, which indicates that the layers are coupled more strongly in CrGeTe$_3$\cite{li2014crxte}. The formation energy of CrSnTe$_3$ is even higher than the other two compounds, revealing that it presents the strongest interlayer coupling, which leads to its 3D characteristics.  Figure~\ref{fig:4}(b), (c), and (d) illustrate the charge density of the three compounds. Consistent with the trend of the $E_\mathrm{f}$ results, the electron density around the Sn-Sn pair is the least among the three materials. Namely, more electrons in CrSnTe$_3$ participate the interlayer coupling.

\section{Conclusions}
In conclusion, we have performed a comprehensive experimental study on the critical properties of single-crystalline CrGeTe$_3$ using isothermal magnetization around the Curie temperature $T_\mathrm{C}$. Based on various experimental techniques including the modified Arrott plot, Kouvel-Fisher method and critical isotherm analysis, we obtained the critical exponents $\beta$, $\gamma$, and $\delta$ of 0.240 $\pm$ 0.006, 1.000 $\pm$ 0.005, and 5.070 $\pm$ 0.006, respectively, at the Curie temperature of 67.9K. These numerical results are similar to the theoretical values in the tricritical mean-field model, which is therefore capable of describing the critical magnetic behavior of 2D CrGeTe$_3$. DFT calculations show that the formation energy of CrGeTe$_3$ lies between those of CrSiTe$_3$ and CrSnTe$_3$, which is in line with a crossover of the magnetic phase transition from 2D to 3D. Overall, our findings provide a fundamental understanding of the anomalous PM-FM transition in a novel 2D ferromagnetic semiconductor.

\begin{acknowledgments}
This work was supported by the National Key Research and Development Program under contracts 2016YFA0300404 and 2016YFA0401003, the Joint Funds of the National Natural Science Foundation of China and the Chinese Academy of Sciences' Large-Scale Scientific Facility under contract U1432139 and U1532153, the National Natural Science Foundation of China under contract 11674326, Key Research Program of Frontier Sciences, CAS (QYZDB-SSW-SLH015), and the Nature Science Foundation of Anhui Province under contract 1508085ME103. J. Z. is supported by the Nature Science Foundation of China (Grant No. 51602079) and the Fundamental Research Funds for the Central Universities of China (Grant No. 372 AUGA5710013115). This research also used computational resources of the National Supercomputing Center of China in Shenzhen (Shenzhen Cloud Computing Center).
\end{acknowledgments}
\bibliography{references}

\begin{thebibliography}{54}%
\makeatletter
\providecommand \@ifxundefined [1]{%
 \@ifx{#1\undefined}
}%
\providecommand \@ifnum [1]{%
 \ifnum #1\expandafter \@firstoftwo
 \else \expandafter \@secondoftwo
 \fi
}%
\providecommand \@ifx [1]{%
 \ifx #1\expandafter \@firstoftwo
 \else \expandafter \@secondoftwo
 \fi
}%
\providecommand \natexlab [1]{#1}%
\providecommand \enquote  [1]{``#1''}%
\providecommand \bibnamefont  [1]{#1}%
\providecommand \bibfnamefont [1]{#1}%
\providecommand \citenamefont [1]{#1}%
\providecommand \href@noop [0]{\@secondoftwo}%
\providecommand \href [0]{\begingroup \@sanitize@url \@href}%
\providecommand \@href[1]{\@@startlink{#1}\@@href}%
\providecommand \@@href[1]{\endgroup#1\@@endlink}%
\providecommand \@sanitize@url [0]{\catcode `\\12\catcode `\$12\catcode
  `\&12\catcode `\#12\catcode `\^12\catcode `\_12\catcode `\%12\relax}%
\providecommand \@@startlink[1]{}%
\providecommand \@@endlink[0]{}%
\providecommand \url  [0]{\begingroup\@sanitize@url \@url }%
\providecommand \@url [1]{\endgroup\@href {#1}{\urlprefix }}%
\providecommand \urlprefix  [0]{URL }%
\providecommand \Eprint [0]{\href }%
\providecommand \doibase [0]{http://dx.doi.org/}%
\providecommand \selectlanguage [0]{\@gobble}%
\providecommand \bibinfo  [0]{\@secondoftwo}%
\providecommand \bibfield  [0]{\@secondoftwo}%
\providecommand \translation [1]{[#1]}%
\providecommand \BibitemOpen [0]{}%
\providecommand \bibitemStop [0]{}%
\providecommand \bibitemNoStop [0]{.\EOS\space}%
\providecommand \EOS [0]{\spacefactor3000\relax}%
\providecommand \BibitemShut  [1]{\csname bibitem#1\endcsname}%
\let\auto@bib@innerbib\@empty
\bibitem [{\citenamefont {Novoselov}\ \emph {et~al.}(2004)\citenamefont
  {Novoselov}, \citenamefont {Geim}, \citenamefont {Morozov}, \citenamefont
  {Jiang}, \citenamefont {Zhang}, \citenamefont {Dubonos}, \citenamefont
  {Grigorieva},\ and\ \citenamefont {Firsov}}]{novoselov2004electric}%
  \BibitemOpen
  \bibfield  {author} {\bibinfo {author} {\bibfnamefont {K.~S.}\ \bibnamefont
  {Novoselov}}, \bibinfo {author} {\bibfnamefont {A.~K.}\ \bibnamefont {Geim}},
  \bibinfo {author} {\bibfnamefont {S.~V.}\ \bibnamefont {Morozov}}, \bibinfo
  {author} {\bibfnamefont {D.}~\bibnamefont {Jiang}}, \bibinfo {author}
  {\bibfnamefont {Y.}~\bibnamefont {Zhang}}, \bibinfo {author} {\bibfnamefont
  {S.~V.}\ \bibnamefont {Dubonos}}, \bibinfo {author} {\bibfnamefont {I.~V.}\
  \bibnamefont {Grigorieva}}, \ and\ \bibinfo {author} {\bibfnamefont {A.~A.}\
  \bibnamefont {Firsov}},\ }\href@noop {} {\bibfield  {journal} {\bibinfo
  {journal} {science}\ }\textbf {\bibinfo {volume} {306}},\ \bibinfo {pages}
  {666} (\bibinfo {year} {2004})}\BibitemShut {NoStop}%
\bibitem [{\citenamefont {Novoselov}\ \emph {et~al.}(2005)\citenamefont
  {Novoselov}, \citenamefont {Geim}, \citenamefont {Morozov}, \citenamefont
  {Jiang}, \citenamefont {Katsnelson}, \citenamefont {Grigorieva},
  \citenamefont {Dubonos},\ and\ \citenamefont {Firsov}}]{novoselov2005two}%
  \BibitemOpen
  \bibfield  {author} {\bibinfo {author} {\bibfnamefont {K.~S.}\ \bibnamefont
  {Novoselov}}, \bibinfo {author} {\bibfnamefont {A.~K.}\ \bibnamefont {Geim}},
  \bibinfo {author} {\bibfnamefont {S.}~\bibnamefont {Morozov}}, \bibinfo
  {author} {\bibfnamefont {D.}~\bibnamefont {Jiang}}, \bibinfo {author}
  {\bibfnamefont {M.}~\bibnamefont {Katsnelson}}, \bibinfo {author}
  {\bibfnamefont {I.}~\bibnamefont {Grigorieva}}, \bibinfo {author}
  {\bibfnamefont {S.}~\bibnamefont {Dubonos}}, \ and\ \bibinfo {author}
  {\bibfnamefont {A.}~\bibnamefont {Firsov}},\ }\href@noop {} {\bibfield
  {journal} {\bibinfo  {journal} {nature}\ }\textbf {\bibinfo {volume} {438}},\
  \bibinfo {pages} {197} (\bibinfo {year} {2005})}\BibitemShut {NoStop}%
\bibitem [{\citenamefont {Zhang}\ \emph {et~al.}(2005)\citenamefont {Zhang},
  \citenamefont {Tan}, \citenamefont {Stormer},\ and\ \citenamefont
  {Kim}}]{zhang2005experimental}%
  \BibitemOpen
  \bibfield  {author} {\bibinfo {author} {\bibfnamefont {Y.}~\bibnamefont
  {Zhang}}, \bibinfo {author} {\bibfnamefont {Y.-W.}\ \bibnamefont {Tan}},
  \bibinfo {author} {\bibfnamefont {H.~L.}\ \bibnamefont {Stormer}}, \ and\
  \bibinfo {author} {\bibfnamefont {P.}~\bibnamefont {Kim}},\ }\href@noop {}
  {\bibfield  {journal} {\bibinfo  {journal} {Nature}\ }\textbf {\bibinfo
  {volume} {438}},\ \bibinfo {pages} {201} (\bibinfo {year}
  {2005})}\BibitemShut {NoStop}%
\bibitem [{\citenamefont {Geim}\ and\ \citenamefont
  {Novoselov}(2007)}]{geim2007rise}%
  \BibitemOpen
  \bibfield  {author} {\bibinfo {author} {\bibfnamefont {A.~K.}\ \bibnamefont
  {Geim}}\ and\ \bibinfo {author} {\bibfnamefont {K.~S.}\ \bibnamefont
  {Novoselov}},\ }\href@noop {} {\bibfield  {journal} {\bibinfo  {journal}
  {Nature materials}\ }\textbf {\bibinfo {volume} {6}},\ \bibinfo {pages} {183}
  (\bibinfo {year} {2007})}\BibitemShut {NoStop}%
\bibitem [{\citenamefont {Castro~Neto}\ \emph {et~al.}(2009)\citenamefont
  {Castro~Neto}, \citenamefont {Guinea}, \citenamefont {Peres}, \citenamefont
  {Novoselov},\ and\ \citenamefont {Geim}}]{RevModPhys.81.109}%
  \BibitemOpen
  \bibfield  {author} {\bibinfo {author} {\bibfnamefont {A.~H.}\ \bibnamefont
  {Castro~Neto}}, \bibinfo {author} {\bibfnamefont {F.}~\bibnamefont {Guinea}},
  \bibinfo {author} {\bibfnamefont {N.~M.~R.}\ \bibnamefont {Peres}}, \bibinfo
  {author} {\bibfnamefont {K.~S.}\ \bibnamefont {Novoselov}}, \ and\ \bibinfo
  {author} {\bibfnamefont {A.~K.}\ \bibnamefont {Geim}},\ }\href {\doibase
  10.1103/RevModPhys.81.109} {\bibfield  {journal} {\bibinfo  {journal} {Rev.
  Mod. Phys.}\ }\textbf {\bibinfo {volume} {81}},\ \bibinfo {pages} {109}
  (\bibinfo {year} {2009})}\BibitemShut {NoStop}%
\bibitem [{\citenamefont {Zibouche}\ \emph {et~al.}(2014)\citenamefont
  {Zibouche}, \citenamefont {Kuc}, \citenamefont {Musfeldt},\ and\
  \citenamefont {Heine}}]{zibouche2014transition}%
  \BibitemOpen
  \bibfield  {author} {\bibinfo {author} {\bibfnamefont {N.}~\bibnamefont
  {Zibouche}}, \bibinfo {author} {\bibfnamefont {A.}~\bibnamefont {Kuc}},
  \bibinfo {author} {\bibfnamefont {J.}~\bibnamefont {Musfeldt}}, \ and\
  \bibinfo {author} {\bibfnamefont {T.}~\bibnamefont {Heine}},\ }\href@noop {}
  {\bibfield  {journal} {\bibinfo  {journal} {Annalen der Physik}\ }\textbf
  {\bibinfo {volume} {526}},\ \bibinfo {pages} {395} (\bibinfo {year}
  {2014})}\BibitemShut {NoStop}%
\bibitem [{\citenamefont {Williams}\ \emph {et~al.}(2015)\citenamefont
  {Williams}, \citenamefont {Aczel}, \citenamefont {Lumsden}, \citenamefont
  {Nagler}, \citenamefont {Stone}, \citenamefont {Yan},\ and\ \citenamefont
  {Mandrus}}]{PhysRevB.92.144404}%
  \BibitemOpen
  \bibfield  {author} {\bibinfo {author} {\bibfnamefont {T.~J.}\ \bibnamefont
  {Williams}}, \bibinfo {author} {\bibfnamefont {A.~A.}\ \bibnamefont {Aczel}},
  \bibinfo {author} {\bibfnamefont {M.~D.}\ \bibnamefont {Lumsden}}, \bibinfo
  {author} {\bibfnamefont {S.~E.}\ \bibnamefont {Nagler}}, \bibinfo {author}
  {\bibfnamefont {M.~B.}\ \bibnamefont {Stone}}, \bibinfo {author}
  {\bibfnamefont {J.-Q.}\ \bibnamefont {Yan}}, \ and\ \bibinfo {author}
  {\bibfnamefont {D.}~\bibnamefont {Mandrus}},\ }\href {\doibase
  10.1103/PhysRevB.92.144404} {\bibfield  {journal} {\bibinfo  {journal} {Phys.
  Rev. B}\ }\textbf {\bibinfo {volume} {92}},\ \bibinfo {pages} {144404}
  (\bibinfo {year} {2015})}\BibitemShut {NoStop}%
\bibitem [{\citenamefont {\ifmmode \check{Z}\else
  \v{Z}\fi{}uti\ifmmode~\acute{c}\else \'{c}\fi{}}\ \emph
  {et~al.}(2004)\citenamefont {\ifmmode \check{Z}\else
  \v{Z}\fi{}uti\ifmmode~\acute{c}\else \'{c}\fi{}}, \citenamefont {Fabian},\
  and\ \citenamefont {Das~Sarma}}]{RevModPhys.76.323}%
  \BibitemOpen
  \bibfield  {author} {\bibinfo {author} {\bibfnamefont {I.}~\bibnamefont
  {\ifmmode \check{Z}\else \v{Z}\fi{}uti\ifmmode~\acute{c}\else \'{c}\fi{}}},
  \bibinfo {author} {\bibfnamefont {J.}~\bibnamefont {Fabian}}, \ and\ \bibinfo
  {author} {\bibfnamefont {S.}~\bibnamefont {Das~Sarma}},\ }\href {\doibase
  10.1103/RevModPhys.76.323} {\bibfield  {journal} {\bibinfo  {journal} {Rev.
  Mod. Phys.}\ }\textbf {\bibinfo {volume} {76}},\ \bibinfo {pages} {323}
  (\bibinfo {year} {2004})}\BibitemShut {NoStop}%
\bibitem [{\citenamefont {MacDonald}\ \emph {et~al.}(2005)\citenamefont
  {MacDonald}, \citenamefont {Schiffer},\ and\ \citenamefont
  {Samarth}}]{macdonald2005ferromagnetic}%
  \BibitemOpen
  \bibfield  {author} {\bibinfo {author} {\bibfnamefont {A.}~\bibnamefont
  {MacDonald}}, \bibinfo {author} {\bibfnamefont {P.}~\bibnamefont {Schiffer}},
  \ and\ \bibinfo {author} {\bibfnamefont {N.}~\bibnamefont {Samarth}},\
  }\href@noop {} {\bibfield  {journal} {\bibinfo  {journal} {Nature Materials}\
  }\textbf {\bibinfo {volume} {4}},\ \bibinfo {pages} {195} (\bibinfo {year}
  {2005})}\BibitemShut {NoStop}%
\bibitem [{\citenamefont {Dietl}(2010)}]{dietl2010ten}%
  \BibitemOpen
  \bibfield  {author} {\bibinfo {author} {\bibfnamefont {T.}~\bibnamefont
  {Dietl}},\ }\href@noop {} {\bibfield  {journal} {\bibinfo  {journal} {Nature
  materials}\ }\textbf {\bibinfo {volume} {9}},\ \bibinfo {pages} {965}
  (\bibinfo {year} {2010})}\BibitemShut {NoStop}%
\bibitem [{\citenamefont {Deng}\ \emph {et~al.}(2011)\citenamefont {Deng},
  \citenamefont {Jin}, \citenamefont {Liu}, \citenamefont {Wang}, \citenamefont
  {Zhu}, \citenamefont {Feng}, \citenamefont {Chen}, \citenamefont {Yu},
  \citenamefont {Arguello}, \citenamefont {Goko} \emph {et~al.}}]{deng2011li}%
  \BibitemOpen
  \bibfield  {author} {\bibinfo {author} {\bibfnamefont {Z.}~\bibnamefont
  {Deng}}, \bibinfo {author} {\bibfnamefont {C.}~\bibnamefont {Jin}}, \bibinfo
  {author} {\bibfnamefont {Q.}~\bibnamefont {Liu}}, \bibinfo {author}
  {\bibfnamefont {X.}~\bibnamefont {Wang}}, \bibinfo {author} {\bibfnamefont
  {J.}~\bibnamefont {Zhu}}, \bibinfo {author} {\bibfnamefont {S.}~\bibnamefont
  {Feng}}, \bibinfo {author} {\bibfnamefont {L.}~\bibnamefont {Chen}}, \bibinfo
  {author} {\bibfnamefont {R.}~\bibnamefont {Yu}}, \bibinfo {author}
  {\bibfnamefont {C.}~\bibnamefont {Arguello}}, \bibinfo {author}
  {\bibfnamefont {T.}~\bibnamefont {Goko}},  \emph {et~al.},\ }\href@noop {}
  {\bibfield  {journal} {\bibinfo  {journal} {Nature communications}\ }\textbf
  {\bibinfo {volume} {2}},\ \bibinfo {pages} {422} (\bibinfo {year}
  {2011})}\BibitemShut {NoStop}%
\bibitem [{\citenamefont {Zhu}\ \emph {et~al.}(2011)\citenamefont {Zhu},
  \citenamefont {Cheng},\ and\ \citenamefont
  {Schwingenschl\"ogl}}]{PhysRevB.84.153402}%
  \BibitemOpen
  \bibfield  {author} {\bibinfo {author} {\bibfnamefont {Z.~Y.}\ \bibnamefont
  {Zhu}}, \bibinfo {author} {\bibfnamefont {Y.~C.}\ \bibnamefont {Cheng}}, \
  and\ \bibinfo {author} {\bibfnamefont {U.}~\bibnamefont
  {Schwingenschl\"ogl}},\ }\href {\doibase 10.1103/PhysRevB.84.153402}
  {\bibfield  {journal} {\bibinfo  {journal} {Phys. Rev. B}\ }\textbf {\bibinfo
  {volume} {84}},\ \bibinfo {pages} {153402} (\bibinfo {year}
  {2011})}\BibitemShut {NoStop}%
\bibitem [{\citenamefont {Xiao}\ \emph {et~al.}(2012)\citenamefont {Xiao},
  \citenamefont {Liu}, \citenamefont {Feng}, \citenamefont {Xu},\ and\
  \citenamefont {Yao}}]{PhysRevLett.108.196802}%
  \BibitemOpen
  \bibfield  {author} {\bibinfo {author} {\bibfnamefont {D.}~\bibnamefont
  {Xiao}}, \bibinfo {author} {\bibfnamefont {G.-B.}\ \bibnamefont {Liu}},
  \bibinfo {author} {\bibfnamefont {W.}~\bibnamefont {Feng}}, \bibinfo {author}
  {\bibfnamefont {X.}~\bibnamefont {Xu}}, \ and\ \bibinfo {author}
  {\bibfnamefont {W.}~\bibnamefont {Yao}},\ }\href {\doibase
  10.1103/PhysRevLett.108.196802} {\bibfield  {journal} {\bibinfo  {journal}
  {Phys. Rev. Lett.}\ }\textbf {\bibinfo {volume} {108}},\ \bibinfo {pages}
  {196802} (\bibinfo {year} {2012})}\BibitemShut {NoStop}%
\bibitem [{\citenamefont {Feng}\ \emph {et~al.}(2012)\citenamefont {Feng},
  \citenamefont {Yao}, \citenamefont {Zhu}, \citenamefont {Zhou}, \citenamefont
  {Yao},\ and\ \citenamefont {Xiao}}]{PhysRevB.86.165108}%
  \BibitemOpen
  \bibfield  {author} {\bibinfo {author} {\bibfnamefont {W.}~\bibnamefont
  {Feng}}, \bibinfo {author} {\bibfnamefont {Y.}~\bibnamefont {Yao}}, \bibinfo
  {author} {\bibfnamefont {W.}~\bibnamefont {Zhu}}, \bibinfo {author}
  {\bibfnamefont {J.}~\bibnamefont {Zhou}}, \bibinfo {author} {\bibfnamefont
  {W.}~\bibnamefont {Yao}}, \ and\ \bibinfo {author} {\bibfnamefont
  {D.}~\bibnamefont {Xiao}},\ }\href {\doibase 10.1103/PhysRevB.86.165108}
  {\bibfield  {journal} {\bibinfo  {journal} {Phys. Rev. B}\ }\textbf {\bibinfo
  {volume} {86}},\ \bibinfo {pages} {165108} (\bibinfo {year}
  {2012})}\BibitemShut {NoStop}%
\bibitem [{\citenamefont {Shan}\ \emph {et~al.}(2013)\citenamefont {Shan},
  \citenamefont {Lu},\ and\ \citenamefont {Xiao}}]{PhysRevB.88.125301}%
  \BibitemOpen
  \bibfield  {author} {\bibinfo {author} {\bibfnamefont {W.-Y.}\ \bibnamefont
  {Shan}}, \bibinfo {author} {\bibfnamefont {H.-Z.}\ \bibnamefont {Lu}}, \ and\
  \bibinfo {author} {\bibfnamefont {D.}~\bibnamefont {Xiao}},\ }\href {\doibase
  10.1103/PhysRevB.88.125301} {\bibfield  {journal} {\bibinfo  {journal} {Phys.
  Rev. B}\ }\textbf {\bibinfo {volume} {88}},\ \bibinfo {pages} {125301}
  (\bibinfo {year} {2013})}\BibitemShut {NoStop}%
\bibitem [{\citenamefont {Xu}\ \emph {et~al.}(2014)\citenamefont {Xu},
  \citenamefont {Yao}, \citenamefont {Xiao},\ and\ \citenamefont
  {Heinz}}]{xu2014spin}%
  \BibitemOpen
  \bibfield  {author} {\bibinfo {author} {\bibfnamefont {X.}~\bibnamefont
  {Xu}}, \bibinfo {author} {\bibfnamefont {W.}~\bibnamefont {Yao}}, \bibinfo
  {author} {\bibfnamefont {D.}~\bibnamefont {Xiao}}, \ and\ \bibinfo {author}
  {\bibfnamefont {T.~F.}\ \bibnamefont {Heinz}},\ }\href@noop {} {\bibfield
  {journal} {\bibinfo  {journal} {Nature Physics}\ }\textbf {\bibinfo {volume}
  {10}},\ \bibinfo {pages} {343} (\bibinfo {year} {2014})}\BibitemShut
  {NoStop}%
\bibitem [{\citenamefont {Han}\ \emph {et~al.}(2014)\citenamefont {Han},
  \citenamefont {Kawakami}, \citenamefont {Gmitra},\ and\ \citenamefont
  {Fabian}}]{han2014graphene}%
  \BibitemOpen
  \bibfield  {author} {\bibinfo {author} {\bibfnamefont {W.}~\bibnamefont
  {Han}}, \bibinfo {author} {\bibfnamefont {R.~K.}\ \bibnamefont {Kawakami}},
  \bibinfo {author} {\bibfnamefont {M.}~\bibnamefont {Gmitra}}, \ and\ \bibinfo
  {author} {\bibfnamefont {J.}~\bibnamefont {Fabian}},\ }\href@noop {}
  {\bibfield  {journal} {\bibinfo  {journal} {Nature nanotechnology}\ }\textbf
  {\bibinfo {volume} {9}},\ \bibinfo {pages} {794} (\bibinfo {year}
  {2014})}\BibitemShut {NoStop}%
\bibitem [{\citenamefont {Ohno}\ \emph {et~al.}(2000)\citenamefont {Ohno},
  \citenamefont {Chiba}, \citenamefont {Matsukura}, \citenamefont {Omiya},
  \citenamefont {Abe}, \citenamefont {Dietl}, \citenamefont {Ohno},\ and\
  \citenamefont {Ohtani}}]{ohno2000electric}%
  \BibitemOpen
  \bibfield  {author} {\bibinfo {author} {\bibfnamefont {H.}~\bibnamefont
  {Ohno}}, \bibinfo {author} {\bibfnamefont {D.}~\bibnamefont {Chiba}},
  \bibinfo {author} {\bibfnamefont {F.}~\bibnamefont {Matsukura}}, \bibinfo
  {author} {\bibfnamefont {T.}~\bibnamefont {Omiya}}, \bibinfo {author}
  {\bibfnamefont {E.}~\bibnamefont {Abe}}, \bibinfo {author} {\bibfnamefont
  {T.}~\bibnamefont {Dietl}}, \bibinfo {author} {\bibfnamefont
  {Y.}~\bibnamefont {Ohno}}, \ and\ \bibinfo {author} {\bibfnamefont
  {K.}~\bibnamefont {Ohtani}},\ }\href@noop {} {\bibfield  {journal} {\bibinfo
  {journal} {Nature}\ }\textbf {\bibinfo {volume} {408}},\ \bibinfo {pages}
  {944} (\bibinfo {year} {2000})}\BibitemShut {NoStop}%
\bibitem [{\citenamefont {Chang}\ \emph {et~al.}(2013)\citenamefont {Chang},
  \citenamefont {Zhang}, \citenamefont {Feng}, \citenamefont {Shen},
  \citenamefont {Zhang}, \citenamefont {Guo}, \citenamefont {Li}, \citenamefont
  {Ou}, \citenamefont {Wei}, \citenamefont {Wang} \emph
  {et~al.}}]{chang2013experimental}%
  \BibitemOpen
  \bibfield  {author} {\bibinfo {author} {\bibfnamefont {C.-Z.}\ \bibnamefont
  {Chang}}, \bibinfo {author} {\bibfnamefont {J.}~\bibnamefont {Zhang}},
  \bibinfo {author} {\bibfnamefont {X.}~\bibnamefont {Feng}}, \bibinfo {author}
  {\bibfnamefont {J.}~\bibnamefont {Shen}}, \bibinfo {author} {\bibfnamefont
  {Z.}~\bibnamefont {Zhang}}, \bibinfo {author} {\bibfnamefont
  {M.}~\bibnamefont {Guo}}, \bibinfo {author} {\bibfnamefont {K.}~\bibnamefont
  {Li}}, \bibinfo {author} {\bibfnamefont {Y.}~\bibnamefont {Ou}}, \bibinfo
  {author} {\bibfnamefont {P.}~\bibnamefont {Wei}}, \bibinfo {author}
  {\bibfnamefont {L.-L.}\ \bibnamefont {Wang}},  \emph {et~al.},\ }\href@noop
  {} {\bibfield  {journal} {\bibinfo  {journal} {Science}\ }\textbf {\bibinfo
  {volume} {340}},\ \bibinfo {pages} {167} (\bibinfo {year}
  {2013})}\BibitemShut {NoStop}%
\bibitem [{\citenamefont {Kou}\ \emph {et~al.}(2014)\citenamefont {Kou},
  \citenamefont {Guo}, \citenamefont {Fan}, \citenamefont {Pan}, \citenamefont
  {Lang}, \citenamefont {Jiang}, \citenamefont {Shao}, \citenamefont {Nie},
  \citenamefont {Murata}, \citenamefont {Tang}, \citenamefont {Wang},
  \citenamefont {He}, \citenamefont {Lee}, \citenamefont {Lee},\ and\
  \citenamefont {Wang}}]{PhysRevLett.113.137201}%
  \BibitemOpen
  \bibfield  {author} {\bibinfo {author} {\bibfnamefont {X.}~\bibnamefont
  {Kou}}, \bibinfo {author} {\bibfnamefont {S.-T.}\ \bibnamefont {Guo}},
  \bibinfo {author} {\bibfnamefont {Y.}~\bibnamefont {Fan}}, \bibinfo {author}
  {\bibfnamefont {L.}~\bibnamefont {Pan}}, \bibinfo {author} {\bibfnamefont
  {M.}~\bibnamefont {Lang}}, \bibinfo {author} {\bibfnamefont {Y.}~\bibnamefont
  {Jiang}}, \bibinfo {author} {\bibfnamefont {Q.}~\bibnamefont {Shao}},
  \bibinfo {author} {\bibfnamefont {T.}~\bibnamefont {Nie}}, \bibinfo {author}
  {\bibfnamefont {K.}~\bibnamefont {Murata}}, \bibinfo {author} {\bibfnamefont
  {J.}~\bibnamefont {Tang}}, \bibinfo {author} {\bibfnamefont {Y.}~\bibnamefont
  {Wang}}, \bibinfo {author} {\bibfnamefont {L.}~\bibnamefont {He}}, \bibinfo
  {author} {\bibfnamefont {T.-K.}\ \bibnamefont {Lee}}, \bibinfo {author}
  {\bibfnamefont {W.-L.}\ \bibnamefont {Lee}}, \ and\ \bibinfo {author}
  {\bibfnamefont {K.~L.}\ \bibnamefont {Wang}},\ }\href {\doibase
  10.1103/PhysRevLett.113.137201} {\bibfield  {journal} {\bibinfo  {journal}
  {Phys. Rev. Lett.}\ }\textbf {\bibinfo {volume} {113}},\ \bibinfo {pages}
  {137201} (\bibinfo {year} {2014})}\BibitemShut {NoStop}%
\bibitem [{\citenamefont {Wang}\ \emph {et~al.}(2015)\citenamefont {Wang},
  \citenamefont {Tang}, \citenamefont {Sachs}, \citenamefont {Barlas},\ and\
  \citenamefont {Shi}}]{PhysRevLett.114.016603}%
  \BibitemOpen
  \bibfield  {author} {\bibinfo {author} {\bibfnamefont {Z.}~\bibnamefont
  {Wang}}, \bibinfo {author} {\bibfnamefont {C.}~\bibnamefont {Tang}}, \bibinfo
  {author} {\bibfnamefont {R.}~\bibnamefont {Sachs}}, \bibinfo {author}
  {\bibfnamefont {Y.}~\bibnamefont {Barlas}}, \ and\ \bibinfo {author}
  {\bibfnamefont {J.}~\bibnamefont {Shi}},\ }\href {\doibase
  10.1103/PhysRevLett.114.016603} {\bibfield  {journal} {\bibinfo  {journal}
  {Phys. Rev. Lett.}\ }\textbf {\bibinfo {volume} {114}},\ \bibinfo {pages}
  {016603} (\bibinfo {year} {2015})}\BibitemShut {NoStop}%
\bibitem [{\citenamefont {Gonz{\'a}lez-Herrero}\ \emph
  {et~al.}(2016)\citenamefont {Gonz{\'a}lez-Herrero}, \citenamefont
  {G{\'o}mez-Rodr{\'\i}guez}, \citenamefont {Mallet}, \citenamefont {Moaied},
  \citenamefont {Palacios}, \citenamefont {Salgado}, \citenamefont {Ugeda},
  \citenamefont {Veuillen}, \citenamefont {Yndurain},\ and\ \citenamefont
  {Brihuega}}]{gonzalez2016atomic}%
  \BibitemOpen
  \bibfield  {author} {\bibinfo {author} {\bibfnamefont {H.}~\bibnamefont
  {Gonz{\'a}lez-Herrero}}, \bibinfo {author} {\bibfnamefont {J.~M.}\
  \bibnamefont {G{\'o}mez-Rodr{\'\i}guez}}, \bibinfo {author} {\bibfnamefont
  {P.}~\bibnamefont {Mallet}}, \bibinfo {author} {\bibfnamefont
  {M.}~\bibnamefont {Moaied}}, \bibinfo {author} {\bibfnamefont {J.~J.}\
  \bibnamefont {Palacios}}, \bibinfo {author} {\bibfnamefont {C.}~\bibnamefont
  {Salgado}}, \bibinfo {author} {\bibfnamefont {M.~M.}\ \bibnamefont {Ugeda}},
  \bibinfo {author} {\bibfnamefont {J.-Y.}\ \bibnamefont {Veuillen}}, \bibinfo
  {author} {\bibfnamefont {F.}~\bibnamefont {Yndurain}}, \ and\ \bibinfo
  {author} {\bibfnamefont {I.}~\bibnamefont {Brihuega}},\ }\href@noop {}
  {\bibfield  {journal} {\bibinfo  {journal} {Science}\ }\textbf {\bibinfo
  {volume} {352}},\ \bibinfo {pages} {437} (\bibinfo {year}
  {2016})}\BibitemShut {NoStop}%
\bibitem [{\citenamefont {Gong}\ \emph {et~al.}()\citenamefont {Gong},
  \citenamefont {Li}, \citenamefont {Li}, \citenamefont {Ji}, \citenamefont
  {Stern}, \citenamefont {Xia}, \citenamefont {Cao}, \citenamefont {Bao},
  \citenamefont {Wang}, \citenamefont {Wang} \emph
  {et~al.}}]{gong2017discovery}%
  \BibitemOpen
  \bibfield  {author} {\bibinfo {author} {\bibfnamefont {C.}~\bibnamefont
  {Gong}}, \bibinfo {author} {\bibfnamefont {L.}~\bibnamefont {Li}}, \bibinfo
  {author} {\bibfnamefont {Z.}~\bibnamefont {Li}}, \bibinfo {author}
  {\bibfnamefont {H.}~\bibnamefont {Ji}}, \bibinfo {author} {\bibfnamefont
  {A.}~\bibnamefont {Stern}}, \bibinfo {author} {\bibfnamefont
  {Y.}~\bibnamefont {Xia}}, \bibinfo {author} {\bibfnamefont {T.}~\bibnamefont
  {Cao}}, \bibinfo {author} {\bibfnamefont {W.}~\bibnamefont {Bao}}, \bibinfo
  {author} {\bibfnamefont {C.}~\bibnamefont {Wang}}, \bibinfo {author}
  {\bibfnamefont {Y.}~\bibnamefont {Wang}},  \emph {et~al.},\ }\href {\doibase
  10.1038/nature22060} {\bibfield  {journal} {\bibinfo  {journal} {Nature}\
  }10.1038/nature22060}\BibitemShut {NoStop}%
\bibitem [{\citenamefont {Sivadas}\ \emph {et~al.}(2015)\citenamefont
  {Sivadas}, \citenamefont {Daniels}, \citenamefont {Swendsen}, \citenamefont
  {Okamoto},\ and\ \citenamefont {Xiao}}]{PhysRevB.91.235425}%
  \BibitemOpen
  \bibfield  {author} {\bibinfo {author} {\bibfnamefont {N.}~\bibnamefont
  {Sivadas}}, \bibinfo {author} {\bibfnamefont {M.~W.}\ \bibnamefont
  {Daniels}}, \bibinfo {author} {\bibfnamefont {R.~H.}\ \bibnamefont
  {Swendsen}}, \bibinfo {author} {\bibfnamefont {S.}~\bibnamefont {Okamoto}}, \
  and\ \bibinfo {author} {\bibfnamefont {D.}~\bibnamefont {Xiao}},\ }\href
  {\doibase 10.1103/PhysRevB.91.235425} {\bibfield  {journal} {\bibinfo
  {journal} {Phys. Rev. B}\ }\textbf {\bibinfo {volume} {91}},\ \bibinfo
  {pages} {235425} (\bibinfo {year} {2015})}\BibitemShut {NoStop}%
\bibitem [{\citenamefont {Carteaux}\ \emph
  {et~al.}(1995{\natexlab{a}})\citenamefont {Carteaux}, \citenamefont {Brunet},
  \citenamefont {Ouvrard},\ and\ \citenamefont
  {Andre}}]{carteaux1995crystallographic}%
  \BibitemOpen
  \bibfield  {author} {\bibinfo {author} {\bibfnamefont {V.}~\bibnamefont
  {Carteaux}}, \bibinfo {author} {\bibfnamefont {D.}~\bibnamefont {Brunet}},
  \bibinfo {author} {\bibfnamefont {G.}~\bibnamefont {Ouvrard}}, \ and\
  \bibinfo {author} {\bibfnamefont {G.}~\bibnamefont {Andre}},\ }\href@noop {}
  {\bibfield  {journal} {\bibinfo  {journal} {Journal of Physics: Condensed
  Matter}\ }\textbf {\bibinfo {volume} {7}},\ \bibinfo {pages} {69} (\bibinfo
  {year} {1995}{\natexlab{a}})}\BibitemShut {NoStop}%
\bibitem [{\citenamefont {Ji}\ \emph {et~al.}(2013)\citenamefont {Ji},
  \citenamefont {Stokes}, \citenamefont {Alegria}, \citenamefont {Blomberg},
  \citenamefont {Tanatar}, \citenamefont {Reijnders}, \citenamefont {Schoop},
  \citenamefont {Liang}, \citenamefont {Prozorov}, \citenamefont {Burch} \emph
  {et~al.}}]{ji2013ferromagnetic}%
  \BibitemOpen
  \bibfield  {author} {\bibinfo {author} {\bibfnamefont {H.}~\bibnamefont
  {Ji}}, \bibinfo {author} {\bibfnamefont {R.}~\bibnamefont {Stokes}}, \bibinfo
  {author} {\bibfnamefont {L.}~\bibnamefont {Alegria}}, \bibinfo {author}
  {\bibfnamefont {E.}~\bibnamefont {Blomberg}}, \bibinfo {author}
  {\bibfnamefont {M.}~\bibnamefont {Tanatar}}, \bibinfo {author} {\bibfnamefont
  {A.}~\bibnamefont {Reijnders}}, \bibinfo {author} {\bibfnamefont
  {L.}~\bibnamefont {Schoop}}, \bibinfo {author} {\bibfnamefont
  {T.}~\bibnamefont {Liang}}, \bibinfo {author} {\bibfnamefont
  {R.}~\bibnamefont {Prozorov}}, \bibinfo {author} {\bibfnamefont
  {K.}~\bibnamefont {Burch}},  \emph {et~al.},\ }\href@noop {} {\bibfield
  {journal} {\bibinfo  {journal} {Journal of Applied Physics}\ }\textbf
  {\bibinfo {volume} {114}},\ \bibinfo {pages} {114907} (\bibinfo {year}
  {2013})}\BibitemShut {NoStop}%
\bibitem [{\citenamefont {Tian}\ \emph {et~al.}(2016)\citenamefont {Tian},
  \citenamefont {Gray}, \citenamefont {Ji}, \citenamefont {Cava},\ and\
  \citenamefont {Burch}}]{tian2016magneto}%
  \BibitemOpen
  \bibfield  {author} {\bibinfo {author} {\bibfnamefont {Y.}~\bibnamefont
  {Tian}}, \bibinfo {author} {\bibfnamefont {M.~J.}\ \bibnamefont {Gray}},
  \bibinfo {author} {\bibfnamefont {H.}~\bibnamefont {Ji}}, \bibinfo {author}
  {\bibfnamefont {R.}~\bibnamefont {Cava}}, \ and\ \bibinfo {author}
  {\bibfnamefont {K.~S.}\ \bibnamefont {Burch}},\ }\href@noop {} {\bibfield
  {journal} {\bibinfo  {journal} {2D Materials}\ }\textbf {\bibinfo {volume}
  {3}},\ \bibinfo {pages} {025035} (\bibinfo {year} {2016})}\BibitemShut
  {NoStop}%
\bibitem [{\citenamefont {Zhuang}\ \emph {et~al.}(2015)\citenamefont {Zhuang},
  \citenamefont {Xie}, \citenamefont {Kent},\ and\ \citenamefont
  {Ganesh}}]{PhysRevB.92.035407}%
  \BibitemOpen
  \bibfield  {author} {\bibinfo {author} {\bibfnamefont {H.~L.}\ \bibnamefont
  {Zhuang}}, \bibinfo {author} {\bibfnamefont {Y.}~\bibnamefont {Xie}},
  \bibinfo {author} {\bibfnamefont {P.~R.~C.}\ \bibnamefont {Kent}}, \ and\
  \bibinfo {author} {\bibfnamefont {P.}~\bibnamefont {Ganesh}},\ }\href
  {\doibase 10.1103/PhysRevB.92.035407} {\bibfield  {journal} {\bibinfo
  {journal} {Phys. Rev. B}\ }\textbf {\bibinfo {volume} {92}},\ \bibinfo
  {pages} {035407} (\bibinfo {year} {2015})}\BibitemShut {NoStop}%
\bibitem [{\citenamefont {Siberchicot}\ \emph {et~al.}(1996)\citenamefont
  {Siberchicot}, \citenamefont {Jobic}, \citenamefont {Carteaux}, \citenamefont
  {Gressier},\ and\ \citenamefont {Ouvrard}}]{siberchicot1996band}%
  \BibitemOpen
  \bibfield  {author} {\bibinfo {author} {\bibfnamefont {B.}~\bibnamefont
  {Siberchicot}}, \bibinfo {author} {\bibfnamefont {S.}~\bibnamefont {Jobic}},
  \bibinfo {author} {\bibfnamefont {V.}~\bibnamefont {Carteaux}}, \bibinfo
  {author} {\bibfnamefont {P.}~\bibnamefont {Gressier}}, \ and\ \bibinfo
  {author} {\bibfnamefont {G.}~\bibnamefont {Ouvrard}},\ }\href@noop {}
  {\bibfield  {journal} {\bibinfo  {journal} {The Journal of Physical
  Chemistry}\ }\textbf {\bibinfo {volume} {100}},\ \bibinfo {pages} {5863}
  (\bibinfo {year} {1996})}\BibitemShut {NoStop}%
\bibitem [{\citenamefont {Leb\`egue}\ \emph {et~al.}(2013)\citenamefont
  {Leb\`egue}, \citenamefont {Bj\"orkman}, \citenamefont {Klintenberg},
  \citenamefont {Nieminen},\ and\ \citenamefont
  {Eriksson}}]{PhysRevX.3.031002}%
  \BibitemOpen
  \bibfield  {author} {\bibinfo {author} {\bibfnamefont {S.}~\bibnamefont
  {Leb\`egue}}, \bibinfo {author} {\bibfnamefont {T.}~\bibnamefont
  {Bj\"orkman}}, \bibinfo {author} {\bibfnamefont {M.}~\bibnamefont
  {Klintenberg}}, \bibinfo {author} {\bibfnamefont {R.~M.}\ \bibnamefont
  {Nieminen}}, \ and\ \bibinfo {author} {\bibfnamefont {O.}~\bibnamefont
  {Eriksson}},\ }\href {\doibase 10.1103/PhysRevX.3.031002} {\bibfield
  {journal} {\bibinfo  {journal} {Phys. Rev. X}\ }\textbf {\bibinfo {volume}
  {3}},\ \bibinfo {pages} {031002} (\bibinfo {year} {2013})}\BibitemShut
  {NoStop}%
\bibitem [{\citenamefont {Li}\ and\ \citenamefont {Yang}(2014)}]{li2014crxte}%
  \BibitemOpen
  \bibfield  {author} {\bibinfo {author} {\bibfnamefont {X.}~\bibnamefont
  {Li}}\ and\ \bibinfo {author} {\bibfnamefont {J.}~\bibnamefont {Yang}},\
  }\href@noop {} {\bibfield  {journal} {\bibinfo  {journal} {Journal of
  Materials Chemistry C}\ }\textbf {\bibinfo {volume} {2}},\ \bibinfo {pages}
  {7071} (\bibinfo {year} {2014})}\BibitemShut {NoStop}%
\bibitem [{\citenamefont {Zhang}\ \emph
  {et~al.}(2015{\natexlab{a}})\citenamefont {Zhang}, \citenamefont {Zhao},
  \citenamefont {Yao},\ and\ \citenamefont {Yang}}]{PhysRevB.92.165418}%
  \BibitemOpen
  \bibfield  {author} {\bibinfo {author} {\bibfnamefont {J.}~\bibnamefont
  {Zhang}}, \bibinfo {author} {\bibfnamefont {B.}~\bibnamefont {Zhao}},
  \bibinfo {author} {\bibfnamefont {Y.}~\bibnamefont {Yao}}, \ and\ \bibinfo
  {author} {\bibfnamefont {Z.}~\bibnamefont {Yang}},\ }\href {\doibase
  10.1103/PhysRevB.92.165418} {\bibfield  {journal} {\bibinfo  {journal} {Phys.
  Rev. B}\ }\textbf {\bibinfo {volume} {92}},\ \bibinfo {pages} {165418}
  (\bibinfo {year} {2015}{\natexlab{a}})}\BibitemShut {NoStop}%
\bibitem [{\citenamefont {Liu}\ \emph {et~al.}(2016{\natexlab{a}})\citenamefont
  {Liu}, \citenamefont {Park}, \citenamefont {Garrity},\ and\ \citenamefont
  {Vanderbilt}}]{PhysRevLett.117.257201}%
  \BibitemOpen
  \bibfield  {author} {\bibinfo {author} {\bibfnamefont {J.}~\bibnamefont
  {Liu}}, \bibinfo {author} {\bibfnamefont {S.~Y.}\ \bibnamefont {Park}},
  \bibinfo {author} {\bibfnamefont {K.~F.}\ \bibnamefont {Garrity}}, \ and\
  \bibinfo {author} {\bibfnamefont {D.}~\bibnamefont {Vanderbilt}},\ }\href
  {\doibase 10.1103/PhysRevLett.117.257201} {\bibfield  {journal} {\bibinfo
  {journal} {Phys. Rev. Lett.}\ }\textbf {\bibinfo {volume} {117}},\ \bibinfo
  {pages} {257201} (\bibinfo {year} {2016}{\natexlab{a}})}\BibitemShut
  {NoStop}%
\bibitem [{\citenamefont {Liu}\ \emph {et~al.}(2016{\natexlab{b}})\citenamefont
  {Liu}, \citenamefont {Zou}, \citenamefont {Zhang}, \citenamefont {Zhou},
  \citenamefont {Wang}, \citenamefont {Wang}, \citenamefont {Qu},\ and\
  \citenamefont {Zhang}}]{liu2016critical}%
  \BibitemOpen
  \bibfield  {author} {\bibinfo {author} {\bibfnamefont {B.}~\bibnamefont
  {Liu}}, \bibinfo {author} {\bibfnamefont {Y.}~\bibnamefont {Zou}}, \bibinfo
  {author} {\bibfnamefont {L.}~\bibnamefont {Zhang}}, \bibinfo {author}
  {\bibfnamefont {S.}~\bibnamefont {Zhou}}, \bibinfo {author} {\bibfnamefont
  {Z.}~\bibnamefont {Wang}}, \bibinfo {author} {\bibfnamefont {W.}~\bibnamefont
  {Wang}}, \bibinfo {author} {\bibfnamefont {Z.}~\bibnamefont {Qu}}, \ and\
  \bibinfo {author} {\bibfnamefont {Y.}~\bibnamefont {Zhang}},\ }\href@noop {}
  {\bibfield  {journal} {\bibinfo  {journal} {Scientific Reports}\ }\textbf
  {\bibinfo {volume} {6}} (\bibinfo {year} {2016}{\natexlab{b}})}\BibitemShut
  {NoStop}%
\bibitem [{\citenamefont {Carteaux}\ \emph
  {et~al.}(1995{\natexlab{b}})\citenamefont {Carteaux}, \citenamefont
  {Moussa},\ and\ \citenamefont {Spiesser}}]{carteaux19952d}%
  \BibitemOpen
  \bibfield  {author} {\bibinfo {author} {\bibfnamefont {V.}~\bibnamefont
  {Carteaux}}, \bibinfo {author} {\bibfnamefont {F.}~\bibnamefont {Moussa}}, \
  and\ \bibinfo {author} {\bibfnamefont {M.}~\bibnamefont {Spiesser}},\
  }\href@noop {} {\bibfield  {journal} {\bibinfo  {journal} {EPL (Europhysics
  Letters)}\ }\textbf {\bibinfo {volume} {29}},\ \bibinfo {pages} {251}
  (\bibinfo {year} {1995}{\natexlab{b}})}\BibitemShut {NoStop}%
\bibitem [{\citenamefont {Kresse}\ and\ \citenamefont
  {Furthm\"uller}(1996)}]{Ref36}%
  \BibitemOpen
  \bibfield  {author} {\bibinfo {author} {\bibfnamefont {G.}~\bibnamefont
  {Kresse}}\ and\ \bibinfo {author} {\bibfnamefont {J.}~\bibnamefont
  {Furthm\"uller}},\ }\href {\doibase 10.1103/PhysRevB.54.11169} {\bibfield
  {journal} {\bibinfo  {journal} {Phys. Rev. B}\ }\textbf {\bibinfo {volume}
  {54}},\ \bibinfo {pages} {11169} (\bibinfo {year} {1996})}\BibitemShut
  {NoStop}%
\bibitem [{\citenamefont {Ceperley}\ and\ \citenamefont {Alder}(1980)}]{Ref37}%
  \BibitemOpen
  \bibfield  {author} {\bibinfo {author} {\bibfnamefont {D.~M.}\ \bibnamefont
  {Ceperley}}\ and\ \bibinfo {author} {\bibfnamefont {B.~J.}\ \bibnamefont
  {Alder}},\ }\href {\doibase 10.1103/PhysRevLett.45.566} {\bibfield  {journal}
  {\bibinfo  {journal} {Phys. Rev. Lett.}\ }\textbf {\bibinfo {volume} {45}},\
  \bibinfo {pages} {566} (\bibinfo {year} {1980})}\BibitemShut {NoStop}%
\bibitem [{\citenamefont {Perdew}\ and\ \citenamefont {Zunger}(1981)}]{Ref38}%
  \BibitemOpen
  \bibfield  {author} {\bibinfo {author} {\bibfnamefont {J.~P.}\ \bibnamefont
  {Perdew}}\ and\ \bibinfo {author} {\bibfnamefont {A.}~\bibnamefont
  {Zunger}},\ }\href {\doibase 10.1103/PhysRevB.23.5048} {\bibfield  {journal}
  {\bibinfo  {journal} {Phys. Rev. B}\ }\textbf {\bibinfo {volume} {23}},\
  \bibinfo {pages} {5048} (\bibinfo {year} {1981})}\BibitemShut {NoStop}%
\bibitem [{\citenamefont {Bl\"ochl}(1994)}]{Ref39}%
  \BibitemOpen
  \bibfield  {author} {\bibinfo {author} {\bibfnamefont {P.~E.}\ \bibnamefont
  {Bl\"ochl}},\ }\href {\doibase 10.1103/PhysRevB.50.17953} {\bibfield
  {journal} {\bibinfo  {journal} {Phys. Rev. B}\ }\textbf {\bibinfo {volume}
  {50}},\ \bibinfo {pages} {17953} (\bibinfo {year} {1994})}\BibitemShut
  {NoStop}%
\bibitem [{\citenamefont {Kresse}\ and\ \citenamefont {Joubert}(1999)}]{Ref40}%
  \BibitemOpen
  \bibfield  {author} {\bibinfo {author} {\bibfnamefont {G.}~\bibnamefont
  {Kresse}}\ and\ \bibinfo {author} {\bibfnamefont {D.}~\bibnamefont
  {Joubert}},\ }\href {\doibase 10.1103/PhysRevB.59.1758} {\bibfield  {journal}
  {\bibinfo  {journal} {Phys. Rev. B}\ }\textbf {\bibinfo {volume} {59}},\
  \bibinfo {pages} {1758} (\bibinfo {year} {1999})}\BibitemShut {NoStop}%
\bibitem [{\citenamefont {Kittel}(2004)}]{Ref41}%
  \BibitemOpen
  \bibfield  {author} {\bibinfo {author} {\bibfnamefont {C.}~\bibnamefont
  {Kittel}},\ }\href {https://books.google.com/books?id=kym4QgAACAAJ} {\emph
  {\bibinfo {title} {Introduction to Solid State Physics}}}\ (\bibinfo
  {publisher} {Wiley},\ \bibinfo {year} {2004})\BibitemShut {NoStop}%
\bibitem [{\citenamefont {Tristan}\ \emph {et~al.}(2005)\citenamefont
  {Tristan}, \citenamefont {Hemberger}, \citenamefont {Krimmel}, \citenamefont
  {Krug~von Nidda}, \citenamefont {Tsurkan},\ and\ \citenamefont
  {Loidl}}]{PhysRevB.72.174404}%
  \BibitemOpen
  \bibfield  {author} {\bibinfo {author} {\bibfnamefont {N.}~\bibnamefont
  {Tristan}}, \bibinfo {author} {\bibfnamefont {J.}~\bibnamefont {Hemberger}},
  \bibinfo {author} {\bibfnamefont {A.}~\bibnamefont {Krimmel}}, \bibinfo
  {author} {\bibfnamefont {H.-A.}\ \bibnamefont {Krug~von Nidda}}, \bibinfo
  {author} {\bibfnamefont {V.}~\bibnamefont {Tsurkan}}, \ and\ \bibinfo
  {author} {\bibfnamefont {A.}~\bibnamefont {Loidl}},\ }\href {\doibase
  10.1103/PhysRevB.72.174404} {\bibfield  {journal} {\bibinfo  {journal} {Phys.
  Rev. B}\ }\textbf {\bibinfo {volume} {72}},\ \bibinfo {pages} {174404}
  (\bibinfo {year} {2005})}\BibitemShut {NoStop}%
\bibitem [{\citenamefont {Arrott}(1957)}]{Ref49}%
  \BibitemOpen
  \bibfield  {author} {\bibinfo {author} {\bibfnamefont {A.}~\bibnamefont
  {Arrott}},\ }\href {\doibase 10.1103/PhysRev.108.1394} {\bibfield  {journal}
  {\bibinfo  {journal} {Phys. Rev.}\ }\textbf {\bibinfo {volume} {108}},\
  \bibinfo {pages} {1394} (\bibinfo {year} {1957})}\BibitemShut {NoStop}%
\bibitem [{\citenamefont {Kaul}(1985)}]{Ref50}%
  \BibitemOpen
  \bibfield  {author} {\bibinfo {author} {\bibfnamefont {S.}~\bibnamefont
  {Kaul}},\ }\href@noop {} {\bibfield  {journal} {\bibinfo  {journal} {Journal
  of magnetism and magnetic materials}\ }\textbf {\bibinfo {volume} {53}},\
  \bibinfo {pages} {5} (\bibinfo {year} {1985})}\BibitemShut {NoStop}%
\bibitem [{\citenamefont {Khan}\ \emph {et~al.}(2012)\citenamefont {Khan},
  \citenamefont {Mandal}, \citenamefont {Mydeen},\ and\ \citenamefont
  {Prabhakaran}}]{Ref51}%
  \BibitemOpen
  \bibfield  {author} {\bibinfo {author} {\bibfnamefont {N.}~\bibnamefont
  {Khan}}, \bibinfo {author} {\bibfnamefont {P.}~\bibnamefont {Mandal}},
  \bibinfo {author} {\bibfnamefont {K.}~\bibnamefont {Mydeen}}, \ and\ \bibinfo
  {author} {\bibfnamefont {D.}~\bibnamefont {Prabhakaran}},\ }\href {\doibase
  10.1103/PhysRevB.85.214419} {\bibfield  {journal} {\bibinfo  {journal} {Phys.
  Rev. B}\ }\textbf {\bibinfo {volume} {85}},\ \bibinfo {pages} {214419}
  (\bibinfo {year} {2012})}\BibitemShut {NoStop}%
\bibitem [{\citenamefont {Banerjee}(1964)}]{Ref52}%
  \BibitemOpen
  \bibfield  {author} {\bibinfo {author} {\bibfnamefont {B.}~\bibnamefont
  {Banerjee}},\ }\href@noop {} {\bibfield  {journal} {\bibinfo  {journal}
  {Physics letters}\ }\textbf {\bibinfo {volume} {12}},\ \bibinfo {pages} {16}
  (\bibinfo {year} {1964})}\BibitemShut {NoStop}%
\bibitem [{\citenamefont {Zhang}\ \emph
  {et~al.}(2015{\natexlab{b}})\citenamefont {Zhang}, \citenamefont {Menzel},
  \citenamefont {Jin}, \citenamefont {Du}, \citenamefont {Ge}, \citenamefont
  {Zhang}, \citenamefont {Pi}, \citenamefont {Tian},\ and\ \citenamefont
  {Zhang}}]{Ref45}%
  \BibitemOpen
  \bibfield  {author} {\bibinfo {author} {\bibfnamefont {L.}~\bibnamefont
  {Zhang}}, \bibinfo {author} {\bibfnamefont {D.}~\bibnamefont {Menzel}},
  \bibinfo {author} {\bibfnamefont {C.}~\bibnamefont {Jin}}, \bibinfo {author}
  {\bibfnamefont {H.}~\bibnamefont {Du}}, \bibinfo {author} {\bibfnamefont
  {M.}~\bibnamefont {Ge}}, \bibinfo {author} {\bibfnamefont {C.}~\bibnamefont
  {Zhang}}, \bibinfo {author} {\bibfnamefont {L.}~\bibnamefont {Pi}}, \bibinfo
  {author} {\bibfnamefont {M.}~\bibnamefont {Tian}}, \ and\ \bibinfo {author}
  {\bibfnamefont {Y.}~\bibnamefont {Zhang}},\ }\href {\doibase
  10.1103/PhysRevB.91.024403} {\bibfield  {journal} {\bibinfo  {journal} {Phys.
  Rev. B}\ }\textbf {\bibinfo {volume} {91}},\ \bibinfo {pages} {024403}
  (\bibinfo {year} {2015}{\natexlab{b}})}\BibitemShut {NoStop}%
\bibitem [{\citenamefont {Fan}\ \emph {et~al.}(2010)\citenamefont {Fan},
  \citenamefont {Ling}, \citenamefont {Hong}, \citenamefont {Zhang},
  \citenamefont {Pi},\ and\ \citenamefont {Zhang}}]{Ref53}%
  \BibitemOpen
  \bibfield  {author} {\bibinfo {author} {\bibfnamefont {J.}~\bibnamefont
  {Fan}}, \bibinfo {author} {\bibfnamefont {L.}~\bibnamefont {Ling}}, \bibinfo
  {author} {\bibfnamefont {B.}~\bibnamefont {Hong}}, \bibinfo {author}
  {\bibfnamefont {L.}~\bibnamefont {Zhang}}, \bibinfo {author} {\bibfnamefont
  {L.}~\bibnamefont {Pi}}, \ and\ \bibinfo {author} {\bibfnamefont
  {Y.}~\bibnamefont {Zhang}},\ }\href {\doibase 10.1103/PhysRevB.81.144426}
  {\bibfield  {journal} {\bibinfo  {journal} {Phys. Rev. B}\ }\textbf {\bibinfo
  {volume} {81}},\ \bibinfo {pages} {144426} (\bibinfo {year}
  {2010})}\BibitemShut {NoStop}%
\bibitem [{\citenamefont {Kouvel}\ and\ \citenamefont {Fisher}(1964)}]{Ref54}%
  \BibitemOpen
  \bibfield  {author} {\bibinfo {author} {\bibfnamefont {J.~S.}\ \bibnamefont
  {Kouvel}}\ and\ \bibinfo {author} {\bibfnamefont {M.~E.}\ \bibnamefont
  {Fisher}},\ }\href {\doibase 10.1103/PhysRev.136.A1626} {\bibfield  {journal}
  {\bibinfo  {journal} {Phys. Rev.}\ }\textbf {\bibinfo {volume} {136}},\
  \bibinfo {pages} {A1626} (\bibinfo {year} {1964})}\BibitemShut {NoStop}%
\bibitem [{\citenamefont {Widom}(1965)}]{Ref55}%
  \BibitemOpen
  \bibfield  {author} {\bibinfo {author} {\bibfnamefont {B.}~\bibnamefont
  {Widom}},\ }\href@noop {} {\bibfield  {journal} {\bibinfo  {journal} {The
  Journal of Chemical Physics}\ }\textbf {\bibinfo {volume} {43}},\ \bibinfo
  {pages} {3898} (\bibinfo {year} {1965})}\BibitemShut {NoStop}%
\bibitem [{\citenamefont {Widom}(1964)}]{Ref56}%
  \BibitemOpen
  \bibfield  {author} {\bibinfo {author} {\bibfnamefont {B.}~\bibnamefont
  {Widom}},\ }\href@noop {} {\bibfield  {journal} {\bibinfo  {journal} {The
  Journal of Chemical Physics}\ }\textbf {\bibinfo {volume} {41}},\ \bibinfo
  {pages} {1633} (\bibinfo {year} {1964})}\BibitemShut {NoStop}%
\bibitem [{\citenamefont {Yang}\ \emph {et~al.}(2007)\citenamefont {Yang},
  \citenamefont {Lee},\ and\ \citenamefont {Li}}]{Ref60}%
  \BibitemOpen
  \bibfield  {author} {\bibinfo {author} {\bibfnamefont {J.}~\bibnamefont
  {Yang}}, \bibinfo {author} {\bibfnamefont {Y.}~\bibnamefont {Lee}}, \ and\
  \bibinfo {author} {\bibfnamefont {Y.}~\bibnamefont {Li}},\ }\href {\doibase
  10.1103/PhysRevB.76.054442} {\bibfield  {journal} {\bibinfo  {journal} {Phys.
  Rev. B}\ }\textbf {\bibinfo {volume} {76}},\ \bibinfo {pages} {054442}
  (\bibinfo {year} {2007})}\BibitemShut {NoStop}%
\bibitem [{\citenamefont {Kim}\ \emph {et~al.}(2002)\citenamefont {Kim},
  \citenamefont {Revaz}, \citenamefont {Zink}, \citenamefont {Hellman},
  \citenamefont {Rhyne},\ and\ \citenamefont {Mitchell}}]{Ref61}%
  \BibitemOpen
  \bibfield  {author} {\bibinfo {author} {\bibfnamefont {D.}~\bibnamefont
  {Kim}}, \bibinfo {author} {\bibfnamefont {B.}~\bibnamefont {Revaz}}, \bibinfo
  {author} {\bibfnamefont {B.~L.}\ \bibnamefont {Zink}}, \bibinfo {author}
  {\bibfnamefont {F.}~\bibnamefont {Hellman}}, \bibinfo {author} {\bibfnamefont
  {J.~J.}\ \bibnamefont {Rhyne}}, \ and\ \bibinfo {author} {\bibfnamefont
  {J.~F.}\ \bibnamefont {Mitchell}},\ }\href {\doibase
  10.1103/PhysRevLett.89.227202} {\bibfield  {journal} {\bibinfo  {journal}
  {Phys. Rev. Lett.}\ }\textbf {\bibinfo {volume} {89}},\ \bibinfo {pages}
  {227202} (\bibinfo {year} {2002})}\BibitemShut {NoStop}%
\bibitem [{\citenamefont {Lin}\ \emph {et~al.}(2016)\citenamefont {Lin},
  \citenamefont {Zhuang}, \citenamefont {Yan}, \citenamefont {Ward},
  \citenamefont {Puretzky}, \citenamefont {Rouleau}, \citenamefont {Gai},
  \citenamefont {Liang}, \citenamefont {Meunier}, \citenamefont {Sumpter} \emph
  {et~al.}}]{Ref62}%
  \BibitemOpen
  \bibfield  {author} {\bibinfo {author} {\bibfnamefont {M.-W.}\ \bibnamefont
  {Lin}}, \bibinfo {author} {\bibfnamefont {H.~L.}\ \bibnamefont {Zhuang}},
  \bibinfo {author} {\bibfnamefont {J.}~\bibnamefont {Yan}}, \bibinfo {author}
  {\bibfnamefont {T.~Z.}\ \bibnamefont {Ward}}, \bibinfo {author}
  {\bibfnamefont {A.~A.}\ \bibnamefont {Puretzky}}, \bibinfo {author}
  {\bibfnamefont {C.~M.}\ \bibnamefont {Rouleau}}, \bibinfo {author}
  {\bibfnamefont {Z.}~\bibnamefont {Gai}}, \bibinfo {author} {\bibfnamefont
  {L.}~\bibnamefont {Liang}}, \bibinfo {author} {\bibfnamefont
  {V.}~\bibnamefont {Meunier}}, \bibinfo {author} {\bibfnamefont {B.~G.}\
  \bibnamefont {Sumpter}},  \emph {et~al.},\ }\href@noop {} {\bibfield
  {journal} {\bibinfo  {journal} {Journal of Materials Chemistry C}\ }\textbf
  {\bibinfo {volume} {4}},\ \bibinfo {pages} {315} (\bibinfo {year}
  {2016})}\BibitemShut {NoStop}%
\end{thebibliography}%
\end{document}